\documentclass{iopart}
\usepackage{iopams}
\usepackage{graphicx}

\def\EW#1{{%
  \ifmmode <\!#1\!>\else $<\!#1\!>$\fi}}
\def\ket#1{{%
  \ifmmode |\,#1\,\rangle \else $|\,#1\,\rangle$\fi}}
\def\bra#1{{%
  \ifmmode \langle\,#1\,| \else $\langle\,#1\,|$\fi}}
\def\Ket#1{{%
  \ifmmode \left|\,#1\,\right\rangle \else $\left|\,#1\,\right\rangle$\fi}}
\def\Bra#1{{%
  \ifmmode \left\langle\,#1\,\right| \else $\left\langle\,#1\,\right|$\fi}}
\def\braket#1#2{{%
  \ifmmode \langle\,#1\,|\,#2\,\rangle \else
  $\langle\,#1\,|\,#2\,\rangle$ \fi}}
\def\braCket#1#2#3{{%
  \ifmmode \langle\,#1\,|\,#2\,|\,#3\,\rangle \else
  $\langle\,#1\,|\,#2\,|\,#3\,\rangle$ \fi}}

\makeatletter
\def\Braket#1#2{%
 \setbox\@tempboxa\hbox{$#1$} \@tempskipa\wd\@tempboxa
 \setbox\@tempboxa\hbox{$#2$}
 \ifdim\ht\@tempboxa<\@tempskipa
   \left\langle\left.\,#1\,\right|\,#2\,\right\rangle
 \else
   \left\langle\,#1\,\left|\,#2\,\right.\right\rangle
 \fi
}
\def\BraCket#1#2#3{  
 \setbox\@tempboxa\hbox{$#1$} \@tempskipa\wd\@tempboxa
 \setbox\@tempboxa\hbox{$#2$} \@tempskipb\wd\@tempboxa
 \setbox\@tempboxa\hbox{$#3$}
 \ifdim\ht\@tempboxa>\@tempskipb
   \ifdim\ht\@tempboxa>\@tempskipa    
     \left\langle\,#1\,\left|\,#2\,\left|\,#3\,\right.\right.\right\rangle
   \else  
     \left\langle\left.\left.\,#1\,\right|\,#2\,\right|\,#3\,\right\rangle
   \fi
 \else
   \ifdim\@tempskipb>\@tempskipa  
     \left\langle\,#1\,\left|\left.\,#2\,\right|\,#3\,\right.\right\rangle
   \else 
     \left\langle\left.\left.\,#1\,\right|\,#2\,\right|\,#3\,\right\rangle
   \fi
 \fi
}
\makeatother

\begin{document}

\jl{9}

\title[Continuous loading of a Ioffe-Pritchard trap]{Continuous loading of cold atoms into a Ioffe-Pritchard magnetic trap}

\author{Piet O Schmidt, Sven Hensler, J\"{o}rg Werner, Thomas Binhammer,
Axel G\"{o}rlitz and Tilman Pfau}

\address{5. Physikalisches Institut, Universit\"{a}t Stuttgart,
  Pfaffenwaldring 57, D-70550 Stuttgart}

\ead{P.Schmidt@physik.uni-stuttgart.de}

\begin{abstract}
We present a robust continuous optical loading scheme for a
Ioffe-Pritchard (IP) type magnetic trap. Atoms are cooled and
trapped in a modified magneto-optical trap (MOT) consisting of a
conventional 2D-MOT in radial direction and an axial molasses. The
radial magnetic field gradient needed for the operation of the
2D-MOT is provided by the IP trap. A small axial curvature and
offset field provide magnetic confinement and suppress spin-flip
losses in the center of the magnetic trap without altering the
performance of the 2D-MOT. Continuous loading of atoms into the IP
trap is provided by radiative leakage from the MOT to a metastable
level which is magnetically trapped and decoupled from the MOT
light. We are able to accumulate 30 times more atoms in the
magnetic trap than in the MOT. The absolute number of $2\times
10^8$~atoms is limited by inelastic collisions. A model based on
rate equations shows good agreement with our data. Our scheme can
also be applied to other atoms with similar level structure like
alkaline earth metals.
\end{abstract}

\submitto{\JOB}


\section{Introduction}

Cold atomic gases have proven to be a useful tool for a wide
variety of fundamental experiments in physics including
Bose-Einstein condensation (BEC) \cite{EFSummer:98} and atom
interferometry \cite{Berman:1997}. The preparation of ultracold
atomic samples for experiments in magnetic traps is a challenging
task involving multi-step cooling and trapping procedures. The
starting point is usually a magneto-optical trap (MOT)
\cite{Townsend:95a}. Transferring the atoms from the MOT into the
magnetic trap (MT) is typically accompanied by a significant loss
in atom number. A crucial point of this transfer process is to
match the size and position of the atomic cloud in the MOT and the
MT (so called "mode matching") to obtain maximum transfer
efficiency and to avoid heating \cite{Ketterle1999a}.

We present a simple scheme for a \textbf{C}ontinuously
\textbf{L}oaded \textbf{I}offe-\textbf{P}ritchard ("CLIP") trap,
which is directly loaded from a magneto-optical trap. The CLIP
trap can serve as a robust source of cold atoms for various atom
optics experiments. It significantly simplifies the preparation of
cold atomic samples in harmonic magnetic traps and can be
implemented for a variety of elements
\cite{Stuhler:2001,Loftus:2002a}. Accumulation of atoms in the
magnetic trap during operation of the MOT removes the need for a
separate transfer step and allows to have more atoms than in the
MOT.

Besides loading a magnetic trap, our scheme could serve as a
source for magnetic waveguide experiments. A fully continuous atom
laser has been on the wish list of many atomic physicists ever
since the successfull formation of a BEC in dilute atomic gases.
In one of the most promising schemes \cite{Mandonnet:2000a}, a
magnetic waveguide is loaded from a MOT and evaporative cooling is
performed on a continuous beam of atoms along the waveguide
transforming the temporal evolution of evaporation into a spatial
evolution. The mode matching of the initial MOT to the magnetic
waveguide is a complicated issue \cite{Dalibard:2001}. Due to its
continuous character, our scheme is an ideal source for this and
other magnetic waveguide and atom interferometry
\cite{Hinds:2001,Andersson:2002} experiments.

The CLIP trap is based on our previously reported continuously
loaded magnetic quadrupole trap \cite{Stuhler:2001}. The new
scheme removes the need for a transfer step from a 3D-quadrupole
into an IP magnetic trap, thus greatly simplifying the preparation
procedure. Another advantage is the possibility to adjust the
aspect ratio in the IP trap, thus reducing the influence of the
dominant density dependent loss mechanisms
\cite{Stuhler:2001,Stuhler:2002}. The CLIP trap employs a modified
magneto-optical trapping scheme allowing us to operate a MOT and a
large volume IP trap overlapped in space and time. Atoms are
magnetically trapped in a long-lived metastable state which is
decoupled from the MOT light. Transfer between MOT and MT is
provided by radiative leakage
\cite{Bell:99,Stuhler:2001,Loftus:2002a} from the excited state,
which is populated by the MOT laser. We have implemented this
scheme with atomic chromium, but also the alkaline earth metals
and e.g. ytterbium are well suited \cite{Loftus:2002a} for that
scheme and several groups are working on an implementation
\cite{Loftus:2002b,Gorlitz:2002a}.

The paper is organized as follows. In Section \ref{scheme} we
present the continuous loading scheme together with a discussion
of the general requirements on the atomic level structure and
possible implementations. In Section \ref{model} we summarize and
extend the rate equation model developed in References
\cite{Stuhler:2001,Stuhler:2002} for the temperature and the
accumulation of magnetically trapped atoms. The experimental setup
and trapping procedure for atomic chromium is described in Section
\ref{setup}. Measurements of the temperature, the number of
trapped atoms and the accumulation efficiency in the CLIP trap are
presented in Section \ref{experiments} and compared to our model.
We conclude with a discussion of possible applications and
extensions of our scheme in Section \ref{conclusion}.

\section{Continuous loading scheme}\label{scheme}
The basic principle of the continuous loading scheme presented
here has been developed and implemented for chromium in a
3D-quadrupole magnetic trap \cite{Stuhler:2001}. In this paper we
show that the scheme can be extended to magnetic traps of the
Ioffe-Pritchard type. We will introduce the concepts of the CLIP
trap with chromium. Requirements and possibilities for the
implementation with other atomic species will be discussed at the
end of this section.
\begin{figure}
    \includegraphics[width=0.6\columnwidth]{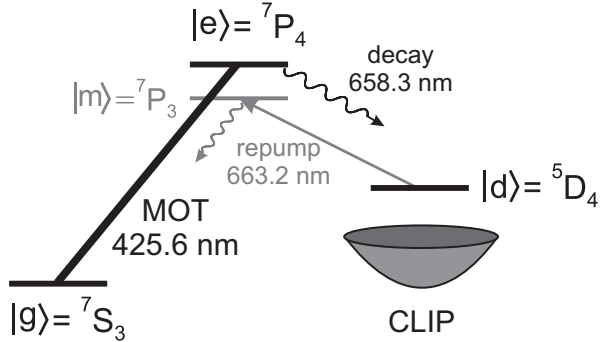}
    \caption{\label{crscheme}CLIP trap loading scheme. Shown are the relevant atomic
    levels and transitions for the implementation
    in $^{52}$Cr (black lines). The magneto-optical trap is operated by driving the fast transition
    from \ket{g} to \ket{e}. Transfer between MOT and CLIP trap is
    provided by radiative leakage from \ket{e} to \ket{d}.
    Repumping the atoms via an intermediate state \ket{m} (gray lines) allows
    to transfer the atoms to the ground state.}
\end{figure}

Figure \ref{crscheme} shows the principle of operation. A strong
dipole transition with linewidth $\Gamma_{eg}$ connecting the
ground state \ket{g} and an excited state \ket{e} allows the
operation of a modified magneto-optical trap. We employ a light
field configuration similar to the 2D$^+$-MOT
\cite{Dieckmann1998a}: two orthogonal pairs of
$\sigma^+/\sigma^-$-polarized laser beams cool and trap the atoms
radially. An additional pair of $\sigma^+$-polarized laser beams
along the axial direction provides Doppler cooling and very weak
confinement due to light pressure forces
\cite{Dalibard:1988,Kazantsev:1989}. For simplicity, in the
following we will refer to our configuration as "MOT", although we
actually mean the modified setup just described. The radial
magnetic field gradient needed for the 2D-MOT is provided by the
Ioffe-Pritchard trap. The magnetic field configuration for the
latter consists in radial direction ($x$, $y$) of a 2D-quadrupole
with magnetic field gradient $B{'}$ supporting the atoms against
gravity. In axial direction ($z$), magnetic confinement is
provided by a curvature field $B{''}$ \cite{Phillips1985a}. Along
the same direction, a small magnetic offset field $B_0$ prevents
Majorana spin-flip losses \cite{Sukumar1997a}.

The cycling transition used for the chromium MOT is not closed.
Atoms can undergo spontaneous emission to a long-lived metastable
state, denoted \ket{d} in Figure \ref{crscheme}, at a rate
$\Gamma_{ed}$. This radiative leakage is the loading mechanism for
the magnetic trap. Atoms in a low field seeking magnetic substate
of the \ket{d}-state manifold can be magnetically trapped in the
field of the IP trap. While operating the MOT, atoms spontaneously
decay to the metastable level and accumulate in the MT. The
branching ratio $\Gamma_{eg}/\Gamma_{ed}$ has to be much larger
than 1 to provide cooling in the MOT before the atoms are
transferred into the MT. In chromium, the branching ratio
$\Gamma_{eg}/\Gamma_{ed}\approx 250000$ is rather high, thus
limiting the loading rate into the CLIP trap (see Section
\ref{modelnum} below). Additional optical pumping to state \ket{m}
would allow a much higher loading rate due to a branching ratio of
only $\Gamma_{mg}/\Gamma_{md}\approx 5200$ and will be implemented
in future experiments. Simultaneous operation of a MOT and a MT is
only possible if the magnetic field gradient required for magnetic
trapping is compatible with the MOT. Chromium with $6~\mu_B$ (Bohr
magnetons) is easily supported against gravity for magnetic field
gradients around 5-10~G/cm, which are typical gradients for
operating a MOT with an atom having a linewidth of $\sim 5$~MHz.

In general, there are few requirements on the atomic species
properties for an implementation of the CLIP trap scheme: (i) The
atoms need to be laser-coolable to operate a MOT. (ii) A
long-lived magnetically trappable state decoupled from the MOT
light must exist. (iii) A dissipative transfer mechanism between
MOT and metastable atoms is needed. (iv) The MOT should operate at
magnetic field parameters required to trap the metastable atoms.

We will briefly discuss each of these points and possible
implementations of the continuous loading scheme for two other
atomic species: rubidium as a representative of the alkali metals
and strontium for the earth alkali metals.

Requirement (i) seems obvious and is met by most atomic species
used in atom optics experiments, although some restrictions may
apply (see below). Besides metastable states, also hyperfine
states with a sufficiently large separation from the states used
for the MOT, should be utilizable to fulfill condition (ii). In
$^{87}$Rb, where the MOT is usually operated on the \ket{g}
$\hat{=}$ $^5$S$_{1/2}$, F=2 $\leftrightarrow$ $^5$P$_3/2$, F=3
$\hat{=}$ \ket{e} transition, the F=1 hyperfine state of the $^5$S
manifold could serve as the magnetic trapping state \ket{m}.
Off-resonant excitation of the $^5$P$_{3/2}$, F=2 state by the MOT
light followed by spontaneous decay to state \ket{m} serves as the
loading mechanism for the magnetic trap (requirement (iii)). The
loading rate can be tuned by adjusting the intensity of the MOT
laser beams. At this point condition (i) puts a restriction on the
actual implementation: a Rb MOT without repumping the atoms from
the F=1 to the F=2 manifold has a strongly reduced efficiency. The
dark SPOT-MOT \cite{Ketterle1993a} solves this problem by
surrounding a central spot in the MOT with repumping light. Also,
the operation of a MOT at magnetic field gradients required for
the magnetic trapping of rubidium having a magnetic moment of
$0.5~\mu_B$ in the lower hyperfine state can be a problem
(requirement (iv)).

A $^{88}$Sr MOT for catching and precooling the atoms is typically
operated between the absolute ground state $^1$S$_0$, \ket{g}, and
the excited state $^1$P$_1^0$. As in chromium, this transition is
not closed. Nevertheless, $8\times 10^7$ atoms can be trapped in a
Sr MOT even without repump laser \cite{Katori:99}. Atoms can
spontaneously decay via an intermediate state $^1$D$_2$ to the
long-lived magnetic trap state $^3$P$_2^0$, \ket{m} (requirements
(ii) and (iii)). Additional decay channels can be closed by
repumping lasers. Here, simultaneous operation of a MOT and a MT
is assisted by a moderate magnetic moment of $3\mu_B$ and the
spectrally broad MOT transition requiring high magnetic field
gradients of 50-150~G/cm by itself (requirement (iv)). A more
elaborate discussion of this trapping scheme and its variants
adopted for ytterbium and other earth alkalis is given in
Reference \cite{Loftus:2002a}.

Requirement (iii) might raise the issue of reabsorption of
spontaneously emitted photons by a very dense cloud of already
magnetically trapped atoms. In the scheme for chromium presented
here, the narrow transition \ket{e}$\leftrightarrow$\ket{d} into
the trap state is spectrally broadened by the strong MOT
transition \ket{g}$\leftrightarrow$\ket{e}. The integral
absorption cross section for this spin-forbidden transition to
state \ket{e} is determined by the small transition strength
giving rise to a suppression of reabsorption by a factor of
$\Gamma_{eg}/\Gamma_{ed}$ \cite{Santos:01a,Stuhler:2002}.

\section{Model}\label{model}
In this section, we want to summarize and extend the model for the
continuous loading scheme developed in Reference
\cite{Stuhler:2001} and adapt it to the Ioffe-Pritchard
configuration.

\subsection{Number of trapped atoms}\label{modelnum}
Loading of the CLIP trap is characterized by a loading rate $R$,
proportional to the number of excited atoms in level \ket{e},
$N^*_\mathrm{MOT}$, the decay rate into the metastable level
$\Gamma_{ed}$, and a transfer efficiency $\eta$, giving
\begin{equation}
\label{loadrate} R=\eta N_\mathrm{MOT}^*\Gamma_{ed}.
\end{equation}
The maximum attainable transfer efficiency (i.e. the fraction of
atoms in a magnetically trappable low field seeking state) can be
estimated using a rate equation model for optical pumping. Its
theoretical prediction for chromium atoms in a standard 3D-MOT is
around 30\% \cite{Stuhler:2002}. Similar values have been obtained
experimentally in our CLIP trap configuration.

Accumulation of atoms in the MT is limited by loss mechanisms
removing magnetically trapped atoms. We have identified two
inelastic collision processes as the major loss mechanisms: (i)
inelastic collisions between optically excited atoms in the MOT
and atoms in the MT, characterized by a rate constant
$\beta_{ed}$, and (ii) inelastic collisions between two
magnetically trapped atoms with a rate constant\footnote{Each
inelastic collision event removes two atoms from the trap.
Therefore the loss rate in the differential equation for the
number of atoms in the trap is twice the inelastic collision
rate.} $\beta_{dd}$. For completeness, we have also included
background gas collisions at a rate $\gamma_d$, although they are
negligible in our setup. The rate equation for the number of atoms
in the CLIP trap, $N_\mathrm{MT}$, reads
\begin{eqnarray}
\label{rateeqn} \frac{\rmd N_\mathrm{MT}}{\rmd t}&=&R-\gamma_d
N_\mathrm{MT} - \beta_{ed} \int\limits_{V}
n_e(\vec{r})n_d(\vec{r})\rmd V -
2\beta_{dd}\int\limits_{V} n^2_d(\vec{r}) \rmd V \nonumber\\
&=&R-(\gamma_d+\gamma_{ed})
N_\mathrm{MT}-2\beta_{dd}\frac{N_\mathrm{MT}^2}{V_\mathrm{MT}}\\
\fl\mathrm{with}\nonumber\\
\label{gammaed}\gamma_{ed}&=&\frac{N_\mathrm{MOT}^*\beta_{ed}}{V_\mathrm{eff}},\\
V_\mathrm{MT}&=& N^2_\mathrm{MT}\left(\int\limits_{V}
n^2_d(\vec{r}) \rmd V\right)^{-1},\\
V_\mathrm{eff}&=& N_\mathrm{MOT}^* N_\mathrm{MT}
\left(\int\limits_{V} n_e(\vec{r})n_d(\vec{r})\rmd V \right)^{-1}.
\end{eqnarray}
Here we have introduced the loss rate $\gamma_{ed}$ for collisions
between MOT and MT atoms to emphasize that this process is
effectively a single-atom loss for magnetically trapped atoms. In
these equations, the density of atoms in the MT and in the excited
state of the MOT is given by $n_d(\vec{r})$ and $n_e(\vec{r})$,
respectively. The size of the effective volume $V_\mathrm{eff}$ is
dominated by the larger of the two volumes $V_\mathrm{MOT}$ and
$V_\mathrm{MT}$ for the magneto-optical and magnetic trap,
respectively. In our experiments with chromium, we find
$V_\mathrm{MOT}\ll V_\mathrm{MT}$, so we can approximate the
effective volume with the volume of the CLIP trap
$V_\mathrm{eff}\approx V_\mathrm{MT}$. Then, the steady state
solution of Equation \ref{rateeqn} is given by
\begin{equation}
\label{steadystate}
 N^\infty_\mathrm{MT}=\frac{-(\gamma_{d}+\gamma_{ed})V_\mathrm{MT}+\sqrt{(\gamma_{d}+\gamma_{ed})^2
 V_\mathrm{MT}^2+8\beta_{dd}RV_\mathrm{MT}}}{2\beta_{dd}}
\end{equation}
Neglecting background gas collisions ($\gamma_d=0$) and assuming
saturation for the MOT transition ($N_\mathrm{MOT}^*\approx
N_\mathrm{MOT}/2$), Equation \ref{steadystate} together with
Equation \ref{gammaed} can be rewritten as an accumulation
efficiency $\kappa$:
\begin{equation}\label{acceff}
\kappa:=\frac{N^\infty_\mathrm{MT}}{N_\mathrm{MOT}}=\frac{-\beta_{ed}+\sqrt{\beta_{ed}^2+16\beta_{dd}RV_\mathrm{MT}/N^2_\mathrm{MOT}}}{4\beta_{dd}}.
\end{equation}
This equation allows an estimate of the number of atoms in the
CLIP trap given the number of atoms in the MOT, $N_\mathrm{MOT}$,
and the inelastic collision properties $\beta_{ed}$ and
$\beta_{dd}$. The inelastic processes limit the achievable density
in the magnetic trap (Equation \ref{rateeqn}) and therefore the
number of accumulated atoms. Increasing the magnetic trapping
volume therefore leads to accumulation of more atoms. This is one
of the main advantages of using the Ioffe-Pritchard trap over a
3D-quadrupole trap: independent control of the trapping field
parameters in radial and axial direction allows increasing the
magnetic trapping volume without degradation of the MOT
performance.

\subsection{Temperature}\label{modeltemp}
The temperature of the atoms in the magnetic trap is determined by
the initial temperature in the MOT ($T_\mathrm{MOT}$) and the
shape of the trapping potentials \cite{Stuhler:2001,Stuhler:2002}.
Assuming the MOT to be much smaller than the MT, loading occurs at
the center of the magnetic trap where the atoms have only kinetic
but no potential energy. In the new trapping potential the initial
kinetic energy is distributed between final kinetic and potential
energy according to the Virial theorem. The trapping potential in
the CLIP trap is harmonic in axial and linear in radial direction,
whereas the MOT is harmonic in all three dimensions. Therefore, in
an ideal situation, one would expect the atoms in the CLIP trap to
have $\frac{1}{2}$ and $\frac{1}{3}$ of the MOT temperature in
axial and radial direction, respectively. Thermalization should
result in a mean final temperature of $\sim 0.4\times
T_\mathrm{MOT}$. In reality additional heating increases the
temperature. One source of heating is misalignment between MOT and
MT due to an imbalance in laser beam intensities. Inelastic
processes like dipolar relaxation collisions \cite{Schmidt:2002b}
and fine-structure changing collisions can also contribute to
heating in the MT.

It is worthwhile mentioning, that it is in fact advantageous to
have the atoms in the MT not too cold if a maximum number of atoms
is desired, since a hot cloud is less dense and thus allows
trapping of more atoms in the presence of density limiting
inelastic losses as discussed in the previous section.

\section{Experimental setup and procedure}\label{setup}
\subsection{Vacuum system and lasers}\label{expsetup}
Our ultra-high vacuum system consists of two vertically arranged
steel chambers connected by a 50~cm long spin-flip Zeeman slower
\cite{Witte:1992a}. With an inner diameter of 1.5~cm, the Zeeman
slower also acts as a differential pumping stage. In the lower
oven-chamber chromium is sublimated at temperatures of around
1700~K in a high temperature effusion cell. A movable metal plate
operated via a mechanical feedthrough allows shutting the atomic
beam on and off within 200~ms. The oven chamber is pumped by a
two-stage turbo-molecular pump assembly to around $10^{-8}$~mbar.
The upper science-chamber is pumped by an ion pump and a
Ti:sublimation pump yielding a pressure in the low $10^{-11}$~mbar
regime. We have implemented a Ioffe-Pritchard trap in the
cloverleaf configuration \cite{Mewes1996a}. Two reentrance vacuum
windows allow the operation of the trap outside the vacuum while
achieving a minimum separation of 3~cm between the cloverleaf coil
assemblies. We achieve a curvature of $B''=116$~G/cm$^2$ in axial
($z$) and a gradient of $B'=150$~G/cm ($x$, $y$) in radial
direction at 300~A current through the coils. All coils are made
of hollow copper tubing and are water cooled during operation.

Cooling and trapping is performed on the
$^7$S$_3\leftrightarrow^7$P$_4$ transition of $^{52}$Cr at a
wavelength of 425.6~nm, a linewidth of $\Gamma_{eg}=2\pi\times
5.02$~MHz and a saturation intensity of
$I_\mathrm{sat}=8.52$~mW/cm$^2$. The laser light for this
transition is generated by frequency doubling the output of an
Ar$^+$-laser pumped Ti:sapphire-laser in an external, pump beam
resonant cavity using a 10~mm long Brewster cut LBO crystal. We
obtain 300~mW of blue light at 1.9~W fundamental input power. The
laser is actively frequency-stabilized to the cooling transition
in the chromium spectrum. Doppler-free polarization spectroscopy
is performed in a cold gas discharge in a chromium tube under an
argon atmosphere. The laser light is split between the Zeeman
slower (120~mW) and the three retroreflected MOT beams (60~mW for
all three beams) with an area of 11~mm$^2$ each. For optimum
performance of the MOT, the $z$ beam was adjusted to have less
than 10~\% of the total MOT intensity. In all experiments reported
here, the laser detuning was set to
$\delta=\omega_\mathrm{laser}-\omega_\mathrm{atom}=-2\Gamma_{eg}$,
unless otherwise noted.

A diode laser at 663.2~nm resonant with the
$^5$D$_4\leftrightarrow ^7$P$_3$ (\ket{d}$\leftrightarrow$\ket{m})
transition provides 5~mW of repumping light in a 2~mm diameter
beam. This transition has a linewidth of $\gamma_{ed}=2\pi\times
127$~1/s that is inhomogeneously broadened by the
$^7$S$_3\leftrightarrow^7$P$_4$ transition. The laser frequency is
locked to a mode of an evacuated and temperature stabilized
Fabry-Perot reference cavity made of Zerodur and Invar.

\subsection{Experimental techniques and data
evaluation}\label{techniques}
We probe the atoms using fluorescence and absorption imaging on
the strong \ket{g}$\leftrightarrow$\ket{e} transition. Before
imaging, magnetically trapped atoms in the metastable \ket{d}
state are optically pumped to the ground state with the repumping
laser. Temperature, shape and magnetic moment of the atoms are not
significantly changed during repumping since only two photons are
scattered by every transferred atom and the involved states
$^7$S$_3$ and $^5$D$_4$ have the same magnetic moment of
$6~\mu_B$. The density profile of the atomic cloud is recorded
with a calibrated CCD camera and fitted with the appropriate
density distribution assuming thermal equilibrium. For the
magneto-optical trap the density distribution is given by a
Gaussian with $1/\sqrt{e}$ radius $\sigma_x=\sigma_y$ and
$\sigma_z$ in radial and axial direction, respectively. The
density in the magnetic trap is given by a Gaussian in axial and
an exponential distribution in radial direction which is modified
by gravity along the $y$-axis:
\begin{eqnarray}
n_\mathrm{MT}(x,y,z)&=&n_0\exp\left(-\frac{\sqrt{x^2+y^2}}{\xi_1}-\frac{y}{\xi_2}-\frac{z^2}{2\sigma^2_{z,\mathrm{MT}}}\right)\\
\fl \mathrm{with}\nonumber\\
\xi_1&=&\frac{k_BT_\mathrm{MT}}{\mu B'}\\
\xi_2&=&\frac{k_BT_\mathrm{MT}}{mg}\\
\sigma_z^\mathrm{MT}&=&\sqrt{\frac{k_BT_\mathrm{MT}}{\mu B''}}.
\end{eqnarray}
In these expressions $n_0$ is the peak density, $k_B$ Boltzmann's
constant,  $g$ the gravitational acceleration and $\mu$, $m$ and
$T_\mathrm{MT}$ are the magnetic moment, the mass and the
temperature of the atoms, respectively. The cloud is imaged along
the $x$-direction onto the CCD camera. For fitting the shape of
the cloud, the density distribution integrated along the imaging
direction is used:
\begin{equation}
n_\mathrm{MT}(y,z)=2n_0\left[\exp\left(-\frac{z^2}{2\sigma^\mathrm{MT}_z}\right)-\frac{y}{\xi_2}|y|\mathrm{K_1}\left(\frac{|y|}{\xi_1}\right)\right],
\end{equation}
where K$_1(x)$ is the modified Bessel function of the second kind
of first order. The number of atoms is determined from the
integrated fluorescence or absorption recorded by the calibrated
CCD camera.

In all our fits, we assume a magnetic moment of 6~$\mu_B$ which is
in most cases a valid approximation since the magnetic trap mainly
supports the extreme Zeeman substate and the change in magnetic
moment during repumping to the ground state is less than 10~\% and
can be safely neglected \cite{Stuhler:2001}. Nevertheless,
accumulation of other than the extreme magnetic substates in
experiments involving high magnetic field gradients or due to
dipolar relaxation processes can lead to some ambiguity in the
determination of the shape of the trapped cloud.

The loading rate into the magnetic trap has been determined from a
measurement in which we loaded the trap for a variable time and
recorded the number of atoms accumulated during that time. A
linear fit to the first 0-250~ms of loading yields the loading
rate. We introduce an effective loading time
$\tau=N_\mathrm{MT}/R$ as a measure for the losses from the trap.
Strictly speaking, this definition is only valid for a
single-particle loss process. Nevertheless $1/\tau$ is a
qualitative measure of the trap loss. More accurately, Equation
\ref{steadystate} should be used. We measured the temperature of
the atomic ensemble by a time-of-flight measurement. In this case,
we also fitted the atoms released from the magnetic trap with a
gaussian density distribution which gave good results after a few
ms time-of-flight.

A typical experimental sequence starts with loading the CLIP trap
from the MOT at a radial gradient of 13~G/cm, an axial curvature
of 11~G/cm$^2$ and an offset field close to zero. After 10~s
loading, MOT and Zeeman-Slower lasers are switched off and the
repumping laser is switched on for 20~ms to transfer the atoms to
the ground state. The magnetic trapping fields are switched off
rapidly ($<300~\mu$s) 11~ms after repumping and - in case of
absorption imaging - a homogeneous magnetic support field along
the imaging axis is switched on. The expanding cloud is imaged
after a variable time-of-flight.
\section{Performance of the CLIP trap}\label{experiments}
In this section, we present measurements on the temperature and
the accumulated number of atoms in the CLIP trap. We compare our
experimental results with the model presented in Section
\ref{model} and give improved numbers for the relevant collisional
properties.
\subsection{Temperature}
\begin{figure}
    \includegraphics[width=0.6\columnwidth]{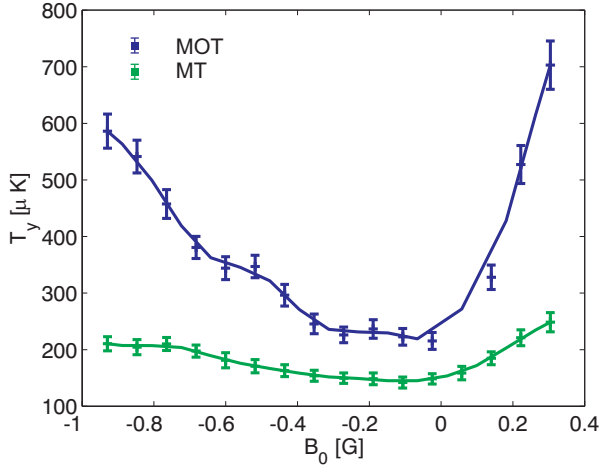}
    \caption{\label{tempyoffs}Radial temperature of the atoms in the
    MOT (black) and MT (gray) for different axial magnetic offset fields.}
\end{figure}

We have measured the temperature of the magneto-optical and
magnetic trap for different light-shift parameters
$(I/I_\mathrm{sat})/(|\delta|/\Gamma_{eg})$. We observed ratios
between the temperature in the MT and in the MOT down to 0.35. For
one set of parameters this can be seen in Figure \ref{tempyoffs}.
This value is in agreement with the ratio predicted from the model
described in Section \ref{modeltemp}. Off-center loading of the
MT, the finite size of the MOT and additional heating mechanisms
(see Section \ref{modelnum}) increase the temperature achieved in
the magnetic trap and therefore lead to a deviation from the model
especially at low temperatures.

One of the major advantages of the Ioffe-Pritchard trap over a
3D-quadrupole trap is the finite offset field at the center of the
trap, preventing atom loss due to Majorana spin-flips for cold
atomic clouds. For the radial trapping parameters in the CLIP
trap, an offset field as low as 40~mG reduces the spin-flip rate
to below 0.1~1/s \cite{Sukumar1997a}. We observed lifetimes of
more than 25~s in the CLIP trap even for negative offset fields.
Although we are not limited by Majorana spin-flips while loading
the CLIP trap, a small positive offset field prevents atom loss
during the subsequent compression of the Ioffe-Pritchard trap.

The influence of an axial magnetic field consisting of the small
offset field and the curvature field of the magnetic trap on the
cooling performance in the MOT is an important aspect
\cite{Walhout:1992,Walhout:1996,Chang:2002a}. In Figure
\ref{tempyoffs} we measured the radial temperature of the MOT and
the MT for different axial magnetic offset field strengths. As
expected, we observe a temperature minimum around zero magnetic
offset field. The number of atoms in the MOT is constant over the
range of magnetic fields shown in Figure \ref{tempyoffs}. For
negative offset fields, the MOT temperature increases only at
large magnetic fields, whereas positive offset fields lead to a
strong degradation of the cooling efficiency of the MOT. For
optimum performance, a magnetic field strength approaching zero
results in a temperature close to the minimum achieveable
temperature in the MOT, while at the same time preventing Majorana
spin-flip losses during subsequent compression of the trap.

In summary, we observed a minimum temperature of $140~\mu$K in the
magneto-optical and $100~\mu$K in the magnetic trap for low MOT
light intensities and large detunings. An axial offset field close
to zero does not degrade the MOT performance. We optimized the
loading of our CLIP trap to achieve a maximum number of
accumulated atoms at the cost of minimal temperature, thus
operating the MOT at high laser light intensities and a detuning
of $-2\Gamma$. Subsequent compression of the IP trap is followed
by a Doppler cooling stage which results in a temperature close to
the Doppler temperature independent of the initial temperature
\cite{Schmidt:2002c}. Using this preparation scheme, the figure of
merit for loading the CLIP trap is reduced to the number of atoms,
thus greatly simplifying adjustment and daily operation.

\subsection{Number of atoms}
In this section, we present experimental results on the number of
atoms accumulated in the CLIP trap as a function of the trap
parameters. We show that the steady state number of atoms is very
robust against moderate magnetic field variations and demonstrate
the advantages of using a Ioffe-Pritchard trap instead of a
3D-quadrupole trap. By determining independently the loading rate
into the CLIP trap and the number of trapped atoms in the MOT and
the MT, we are able to explain our findings qualitatively. A fit
of the model developed in Section \ref{modelnum} to the
experimental data results in a more accurate determination of the
collision parameters responsible for trap loss than previously
reported \cite{Stuhler:2001}.

\begin{figure}
    \includegraphics[width=0.6\columnwidth]{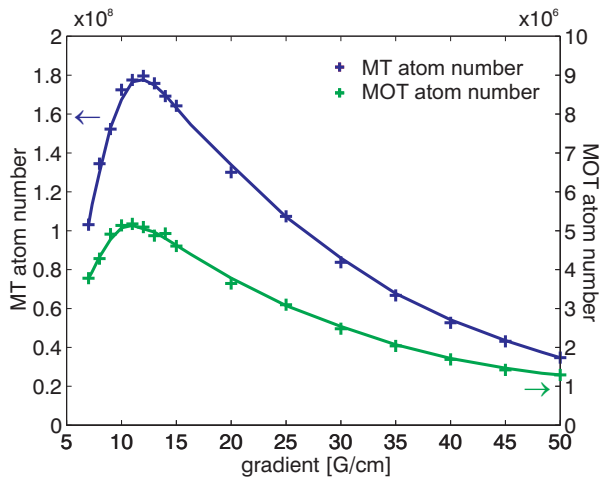}
    \caption{\label{gradnum}Steady state number of atoms in the MOT (gray) and in the CLIP trap
    (black) for different radial magnetic field gradients.
    The lines are 3-point moving averages of the data to guide the eye.
    Note the different scale for the number of atoms in the MOT
    and in the MT.}
\end{figure}
\begin{figure}
    \includegraphics[width=0.6\columnwidth]{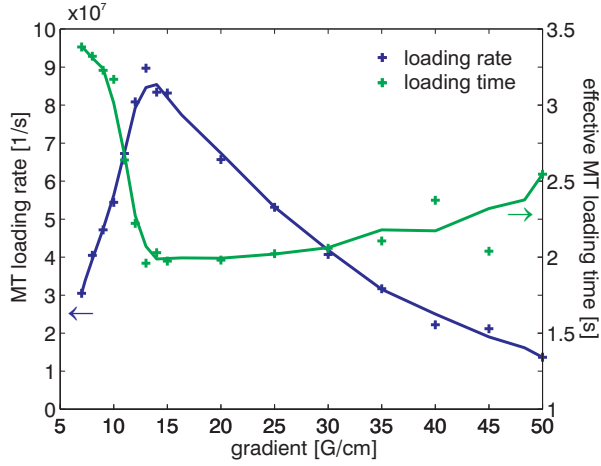}
    \caption{\label{gradload}Loading rate into the CLIP trap (black)
    and effective loading time (gray) for different radial magnetic field
    gradients.
    The lines are 3-point moving averages of the data to guide the eye.}
\end{figure}
In Figure \ref{gradnum}, we have plotted the number of atoms in
the MOT and the steady state number of atoms accumulated in the
CLIP trap for various radial field gradients while keeping all
other trapping parameters at their optimum values. The measured
volume of the magnetic trap decreases almost linearly with
increasing gradient from $14\times 10^{-3}$~cm$^3$ to $4\times
10^{-3}$~cm$^3$. Due to an unusually high inelastic excited state
collision rate for chromium \cite{Bradley:00a, Stuhler:2002}, the
number of atoms in the MOT is density limited to around $5\times
10^6$. The number of atoms in the MT exhibits a maximum around
13~G/cm. To either side, a decrease in atom number coincides with
a reduced MOT performance and thus a reduced loading rate (see
Equation \ref{loadrate}). This situation is comparable to changing
the gradient of a 3D-quadrupole trap \cite{Stuhler:2001}: both,
magnetic and magneto-optical trap are affected by a change in
gradient. This can be seen more clearly in the loading rate and
time measurement presented in Figure \ref{gradload}. The loading
rate closely follows the number of atoms in the MOT for high
gradients and decreases even more steeply for low gradients. This
behaviour is an indication for a degradation in loading efficiency
due to position mismatch between MOT and MT. The inverse of the
effective loading time $\tau$, which takes into account inelastic
collisions, is a measure for the loss rate from the magnetic trap
(see Section \ref{techniques}). Therefore the observed increase in
effective loading time for low gradients in Figure \ref{gradload}
resembles a decreased loss from the CLIP trap originating from a
reduced number of atoms in the MOT and an increase of the trapping
volume (see Equations \ref{rateeqn} and \ref{gammaed}).

For high gradient fields, we observe a small rise in the effective
loading time. At this point the number of atoms in the magnetic
trap decreases faster than the trapping volume, thus the density
in the CLIP trap decreases. This leads to a slight reduction in
the inelastic loss rates and therefore an increase in loading
time.

\begin{figure}
    \includegraphics[width=0.6\columnwidth]{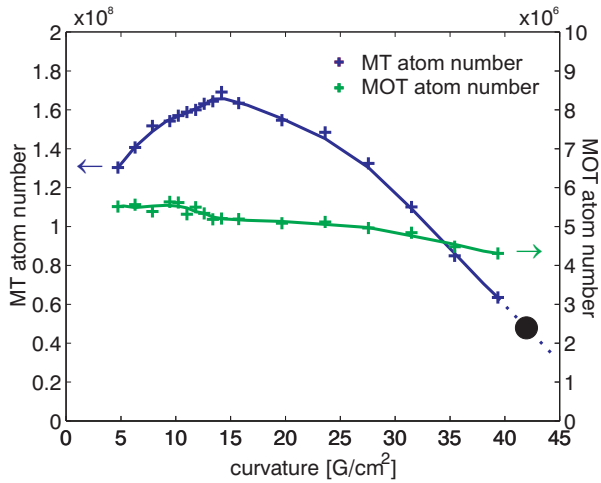}
    \caption{\label{curvnum}Steady state number of atoms in the MOT (gray) and in the CLIP trap
    (black) for different axial magnetic field curvatures.
    The filled black circle shows a typical 3D-quadrupole trap having a
    volume that corresponds to
    the volume of an IP trap with the indicated curvature.
    The lines are 3-point moving averages of the data to guide the
    eye.
    Note that the vertical scales are identical to Figure \ref{gradnum}.}
\end{figure}
\begin{figure}
    \includegraphics[width=0.6\columnwidth]{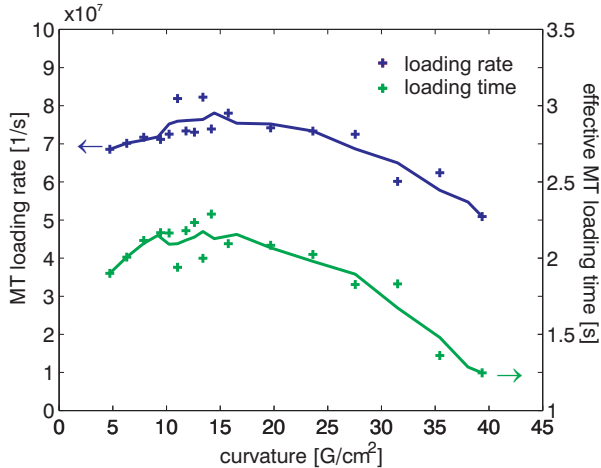}
    \caption{\label{curvload}Loading rate into the CLIP trap (black)
    and effective loading time (gray) for different axial magnetic field curvatures.
    The lines are 3-point moving averages of the data to guide the eye.
    Note that the vertical scales are identical to Figure \ref{gradload}.}
\end{figure}
The situation shown in Figures \ref{curvnum} and \ref{curvload} is
different. Here, we have changed the axial curvature and recorded
the number of atoms in the MOT and the CLIP trap as well as the
effective loading time and rate. The volume of the magnetic trap
is approximately inversely proportional to the applied curvature
field, ranging from $14\times 10^{-3}$ to $3.9\times 10^{-3}$
~cm$^3$ thus covering essentially the same range as for the
gradient measurements.  For comparison, we have marked in Figure
\ref{curvnum} the equivalent curvature of a typical 3D-quadrupole
trap having a volume corresponding to the volume in the CLIP trap.
Whereas in Figure \ref{gradnum} the MOT performance is strongly
affected by the radial gradient, in Figure \ref{curvnum} only a
weak dependence of the number of atoms in the MOT on the curvature
field is observed (note that the vertical axis scales in both
Figures are identical). Consequently, the loading rate in Figure
\ref{curvload} is also constant to within 30~\%. In this
situation, the number of atoms accumulated in the CLIP trap is
limited mainly by inelastic loss processes. This becomes evident
from the decrease in the effective loading time (i.e. the increase
in loss) in the high curvature regime shown in Figure
\ref{curvload}. At very low curvatures reduced axial confinement,
apparent in the reduced effective loading time, and probably
misalignment between MOT and MT, noticeable in a slightly reduced
loading rate, are responsible for a lower number of atoms in the
CLIP trap. More accurate alignment of the MOT at low curvature
fields should allow a constant loading rate at its maximum value
even at low curvature fields. Such a configuration would be useful
for continuous loading of a magnetic waveguide.

We achieved a maximum number of $2\times 10^8$ atoms in the CLIP
trap at a loading rate of $10^8$ atoms/s for optimum trapping
parameters of $B{'}=12.5$~G/cm for the radial gradient,
$B{''}=10.5$~G/cm$^2$ for the axial curvature and an axial offset
field close to zero. This is about three times the number of atoms
trapped previously in a 3D-quadrupole trap under comparable
loading conditions.

\begin{figure}
    \includegraphics[width=0.6\columnwidth]{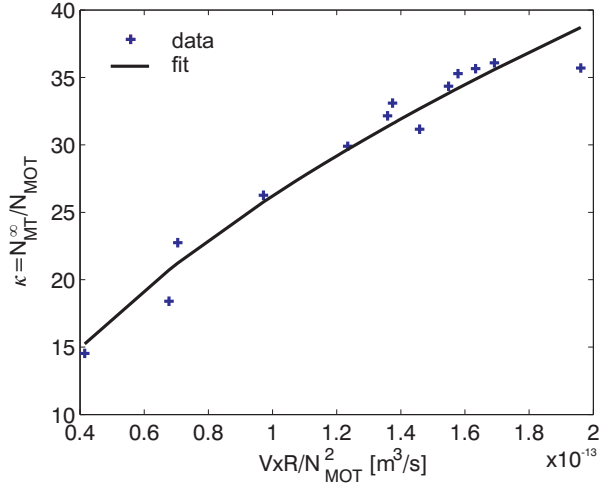}
    \caption{\label{acceffplot}Accumulation efficiency into the CLIP trap
    versus transfer rate divided by the density of excited MOT atoms.
    The line is a least square fit to the data.}
\end{figure}
The data presented in Figures \ref{gradload} and \ref{curvload}
indicates that the effective loading time is limited to around 2~s
due to inelastic collisions. We have determined the magnitude of
the two major loss mechanisms by fitting the simplified rate
equation model from Equation \ref{acceff} to the gradient and
curvature data from Figures \ref{gradnum} and \ref{curvnum}. The
result is shown in Figure \ref{acceffplot}, where we have plotted
the accumulation efficiency $\kappa=N_\mathrm{MT}/N_\mathrm{MOT}$
versus $RV_\mathrm{MT}/N_\mathrm{MOT}^{2}$. The line is a least
square fit of Equation \ref{acceff} to the data with $\beta_{dd}$
and $\beta_{ed}$ as fitting parameters. For the fit we have
neglected the low gradient and curvature data points in which
other loss mechanisms dominate. Since both loss rates scale with
the volume of the magnetic trap, the fitting parameters are
strongly correlated resulting in the large statistical uncertainty
for each parameter. We obtain $\beta_{ed}=6(9)\times
10^{-10}$~cm$^3$/s for inelastic collisions between MOT atoms in
the excited state and magnetically trapped atoms and
$\beta_{dd}=1.3(5)\times 10^{-11}$~cm$^3$/s for inelastic
collisions between atoms in the CLIP trap. Systematic errors,
mainly from the determination of the densities, reduce the
accuracy of the given values to within a factor of 2. These more
accurate results improve the order of magnitude rates reported
previously \cite{Stuhler:2001}. An independent measurement of
$\beta_{dd}$ was performed by observing the decay of magnetically
trapped atoms in the $^5$D$_4$ state and fitting a rate equation
including single- and two-atom loss processes to the data. The
experimental situation in the decay measurement is different from
the accumulation measurement, where only the steady state atom
number is recorded. Other processes not included in our model,
like e.g. dipolar relaxation \cite{Schmidt:2002b} and variations
in temperature and magnetic substate distribution of the cloud
during the decay process make a comparison difficult.
Nevertheless, the obtained value of $\beta_{dd}=3.8(4)\times
10^{-11}$~cm$^3$/s is close to the value from the fit to the
accumulation efficiency.

\section{Conclusion}\label{conclusion}
We have presented a continuous optical loading scheme for
ultracold atoms from a magneto-optical trap into a Ioffe-Pritchard
magnetic trap. We achieved temperatures below $100~\mu$K, a
loading rate of up to $10^8$ atoms/s and $2\times 10^8$ atoms
accumulated in the magnetic trap in our implementation for
chromium atoms. The loading rate is limited by the small number of
atoms trappable in a chromium MOT \cite{Bradley:00a,Stuhler:2002}
and the small decay rate $\Gamma_{ed}$ into the metastable trap
state. We plan to increase the loading rate by optically pumping
the atoms via another fine structure level of the excited state
that has a higher decay rate into the trap state. Two major loss
mechanisms limit the number of atoms in the magnetic trap:
inelastic collisions among excited state atoms in the MOT and
magnetically trapped atoms and inelastic collisions between atoms
in the magnetic trap \cite{Stuhler:2001}. We have presented a
model for the steady state number of atoms in the magnetic trap
that is in good agreement with our experimental data. From a fit
of the model to the data we could determine the loss rates for the
two inelastic density limiting processes to be $\beta_{ed}=5\times
10^{-10}\pm 45~\%$~cm$^3$/s and $\beta_{dd}=1.3\times 10^{-11}\pm
17~\%$~cm$^3$/s for collisions between MOT and MT atoms and
between MT atoms, respectively. Independent control of the radial
and axial trapping fields in the Ioffe-Pritchard trap allowed us
to accumulate more atoms in the MT by increasing the volume of the
trap without losing confinement or deteriorating MOT performance.
The inelastic loss in the metastable trap state in chromium
requires repumping the atoms to the ground state for subsequent
experiments.

We use the CLIP trap loading scheme as a starting point for
further cooling sequences. Consecutive Doppler cooling of ground
state atoms in a compressed IP magnetic trap reduces the figure of
merit for loading of the MT to the number of atoms instead of
phase space density \cite{Schmidt:2002c}. Radio frequency induced
evaporative cooling towards a Bose-Einstein Condensate of chromium
atoms is currently under investigation.

Our experiments at low axial confinement (Figures \ref{curvnum},
\ref{curvload}) show that continuous loading of a magnetic
waveguide should be possible. In that case, slightly tilting the
trap would allow the atoms to escape the trapping region resulting
in a continuous flux of $10^8$ magnetically trapped ultracold
atoms per second in the case of chromium.

Another possibility of accumulating orders of magnitudes more
atoms would be to extend the Ioffe-Pritchard configuration. For
example the combination of a tilted magnetic waveguide with
magnetic or optical endcaps could serve as a large volume
accumulation reservoir and prolong the loading time.

The continuously loaded Ioffe-Pritchard trap presented here is not
limited to chromium. Atoms like e.g. the earth alkalis and
ytterbium with a large natural linewidth allow high gradients for
operating the MOT, thus enabling magnetic trapping at the same
time. Besides this feature, the earth alkalis have a level
structure which is especially well suited for the continuous
loading scheme presented here or variations of it
\cite{Loftus:2002a}.

\ack This work was funded by the Forschergruppe "Quantengase" der
Deutschen Forschungsgemeinschaft and the European Research and
Training Network "Cold Quantum Gases" under Contract No.
HPRN-CT-2000-00125. P.O.S has been supported by the
Studien\-stiftung des deutschen Volkes.


\begin{thebibliography}{10}
\bibitem{Gorlitz:2002a}
We plan to implement this loading scheme for a different
experiment with
  ytterbium atoms.
\end{thebibliography}
\end{document}